\documentclass[pra,showpacs,amsmath,amssymb, twocolumn, 10pt]{revtex4}

\usepackage{amsthm}
\usepackage{dcolumn}
\usepackage{bm}
\usepackage{graphicx}

\begin{document}
\newtheorem{corollary}{Corollary}
\newtheorem{definition}{Definition}
\newtheorem{example}{Example}
\newtheorem{lemma}{Lemma}
\newtheorem{proposition}{Proposition}
\newtheorem{theorem}{Theorem}
\newtheorem{fact}{Fact}
\newtheorem{property}{Property}
\newcommand{\bra}[1]{\langle #1|}
\newcommand{\ket}[1]{|#1\rangle}
\newcommand{\braket}[3]{\langle #1|#2|#3\rangle}
\newcommand{\ip}[2]{\langle #1|#2\rangle}
\newcommand{\op}[2]{|#1\rangle \langle #2|}

\newcommand{\tr}{{\rm tr}}
\newcommand{\supp}{{\it supp}}
\newcommand{\sch}{{\it Sch}}

\newcommand {\E } {{\mathcal{E}}}
\newcommand {\F } {{\mathcal{F}}}
\newcommand {\diag } {{\rm diag}}

\title{Conditions for entanglement transformation between a class of multipartite pure states with generalized Schmidt decompositions}
\author{Yu Xin$^{1,2}$}
\email{xiny05@mails.tsinghua.edu.cn}
\author{Runyao Duan$^{1}$}
\email{dry@tsinghua.edu.cn}

\affiliation{$^1$State Key Laboratory of Intelligent Technology and
Systems, Department of Computer Science and Technology, Tsinghua
University, Beijing 100084, China,
\\
$^2$Department of Physics, Tsinghua University, Beijing 100084,
China}

\date{\today}

\begin{abstract}
We generalize  Nielsen's marjoization criterion for the
convertibility of bipartite pure states [Phys. Rev. Lett
\textbf{83}, 436(1999)] to a special class of multipartite pure
states with generalized Schmidt decompositions.
\end{abstract}


\maketitle

One of the central problems of quantum entanglement is to find
conditions under which  an entangled state can be transformed into
another one by local operations and classical communication
\cite{BBPS96}. In 1999 Nielsen reported a sufficient and necessary
condition for the deterministic entanglement transformations between
bipartite pure states \cite{NIE99}. Nielsen's work has been extended
to the case when a deterministic transformation cannot be achieved
\cite{LO97, JP99a, JP99, VID99, VJN00, FM00,SRS02, FWX02, DK01,
FDY04, DFY05, SDY05,CS05,DJFY06,AN07,TUR07,OMM04}. These efforts
indicate that the structure of bipartite pure entanglement has been
well understood.

But all the works above are only on bipartite pure states and very
little is known about the structure of multipartite pure states
\cite{BPRS+01}. Recently the study of entanglement transformation
between multipartite states has received considerable attentions
\cite{SVW05, CC06}. It is of great interest to generalize Nielsen's
result to a multipartite scenario. Such a result should lead to a
deep understanding of the nature of multipartite entanglement. For
instance, it can be used to  identify what kind of multipartite
entangled states are universal resources in realizing one-way
quantum computation \cite{RB01}.

In this Brief Report, we consider entanglement transformations of a
class of multipartite pure states which have generalized Schmidt
decompositions. We show that Nielsen's theorem can be extended to
this class of states. Furthermore, our result confirms the intuition
that the entanglement of a multipartite state with a generalized
Schmidt decomposition is completely determined by its Schmidt
coefficient vector.

Suppose Alice, Bob, ..., and Dana share a multipartite pure state
$\ket{\psi}$ which has a generalized Schmidt decomposition as
follows:
\begin{equation}\label{source}
\ket{\psi}=\sum_{k=0}^{n-1} \sqrt{\lambda_k}\ket{k}_A\ket{k}_B\cdots
\ket{k}_D,
\end{equation}
where $\{\ket{k}\}_A$, $\{\ket{k}\}_B$, ..., and $\{\ket{k}\}_D$ are
orthonormal bases for Alice, Bob,..., and Dana respectively, and
$\lambda=(\lambda_0,\ldots, \lambda_{n-1})$ represents the Schmidt
coefficient vector with nonincreasing order$-$ i.e., $\lambda_0\geq
\cdots\geq \lambda_{n-1}\geq 0$ . They want to transform
$\ket{\psi}$ to the following state $\ket{\phi}$ using LOCC:
\begin{equation}\label{target}
\ket{\phi}=\sum_{k=0}^{n-1} \sqrt{\mu_k}\ket{k'}_A\ket{k'}_B\cdots
\ket{k'}_D,
\end{equation}
where $\{\ket{k'}\}_A$, $\{\ket{k'}\}_B$, ..., and $\{\ket{k'}\}_D$
are also orthonormal bases for Alice, Bob,..., and Danna,
respectively, and $\mu=(\mu_0,\ldots, \mu_{n-1})$ is the Schmidt
coefficients vector with non-increasing order.  Two sets of bases
$\{\ket{k}\}$ and $\{\ket{k'}\}$ are generally different. We say
$\lambda$ is majorized by $\mu$, denoted as $\lambda\prec \mu$, if
$$\sum_{k=0}^l \lambda_k\leq \sum_{k=0}^l \mu_k,~0\leq l\leq n-2,$$ and
$\sum_{k=0}^{n-1}\lambda_k=\sum_{k=0}^{n-1}\mu_k$.  With these
notations, we have the following generalization of Nielsen's
theorem.
\begin{theorem}\label{schmidtsep}\upshape
Alice, Bob,..., Dana can transform $\ket{\psi}$ to $\ket{\phi}$
using LOCC if and only if $\lambda\prec\mu$.
\end{theorem}

When there are only two parties, the above theorem is exactly the
one by Nielsen \cite{NIE99}. This is due to the fact that any
bipartite pure state has a Schmidt decomposition. In general, there
exist multipartite states not having such a decomposition
\cite{PER95}. So the above theorem only covers a special class of
multipartite states. We would like to point out that a
characterization of the existence of a generalized Schmidt
decomposition has already been given by Thapliyal \cite{THA99}.

Since we can split all these parties into two groups and treat each
group altogether as a single party, the necessity of the condition
$\lambda\prec\mu$ follows directly from Nielsen's theorem. However,
whether this condition is also sufficient for the convertibility
remains unknown. Actually, the protocol given by Nielsen consists of
a local measurement by one party and conditional unitary operations
by the other party. When there are more than two parties, it is not
clear whether all these operations can be implemented locally.
Fortunately, we notice that in Ref. \cite{JS01} Jensen and Schack
have presented an alternative protocol for entanglement
transformations between bipartite pure states, which simplifies the
Nielsen's original proof considerably. Although they are only
concerned with bipartite pure states, we shall show that with minor
modifications their protocol can be used to prove that majorization
is also a sufficient condition for entanglement transformations
between multipartite states with generalized Schmidt decompositions.

We shall employ a useful alternative characterization of
majorization \cite{MO79}. That is, if $\lambda\prec \mu$, then we
can write $\lambda=\sum_{j=1}^{N}p_j\sigma_j\mu$ for a probability
distribution $\{p_j\}$ and a set of permutations $\{\sigma_j\}$,
where $N$ is at most $(n-1)^2+1$.  More explicitly, we have
\begin{equation}\label{keyeq}
\sum_{j=1}^N p_{j}\mu_{\sigma^{-1}_j(k)}=\lambda_k,~~k=0,\ldots,
n-1.
\end{equation}

Let
\begin{equation}\label{keyeq2}
M^A_j=\sqrt{p_j}\sum_{k=0}^{n-1}\sqrt{\frac{\mu_{\sigma^{-1}_j(k)}}{\lambda_k}}\op{k_A}{k_A},~~j=1,\ldots,
N.
\end{equation}

It follows from Eq. (\ref{keyeq}) that $\sum_{j=1}^N {M^A_j}^\dagger
M^A_j=I^A$. So $\{M^A_j\}$ is a complete quantum measurement. A
simple local protocol that transforms $\ket{\psi}$ to $\ket{\phi}$
consists of the following two steps.

Step 1. Alice performs $\{M^A_j\}$ on her subsystem, then broadcasts
the measurement outcome $j$ to other parties;

Step 2. Every party performs a unitary operation $U_j$ on his/her
own subsystem if the measurement outcome is $j$, where
$U_j=\sum_{k=0}^{n-1} \op{\sigma^{-1}_j(k)'}{k}.$

The validity of the above protocol can be verified as follows. By
Eqs. (\ref{keyeq}) and (\ref{keyeq2}), Alice obtains the measurement
outcome $j$ with probability $p_j$, and the post-measurement state
is changed into
\begin{equation}\label{source}
\ket{\psi_j}=\sum_{k=0}^{n-1}
\sqrt{\mu_{\sigma^{-1}_j(k)}}\ket{k}_A\ket{k}_B\cdots \ket{k}_D.
\end{equation}
Then every party then performs a unitary operation $U_j$ on his/her
own subsystem. After that, $\ket{\psi_j}$ is transformed into
\begin{equation}\label{source}
\ket{\phi_j}=\sum_{k=0}^{n-1}\sqrt{\mu_{\sigma^{-1}_j(k)}}\ket{\sigma^{-1}_j(k)'}_A\ket{\sigma^{-1}_j(k)'}_B\cdots
\ket{\sigma^{-1}_j(k)'}_D.
\end{equation}
Relabeling  the subscript $\sigma^{-1}_j(k)$ as $k$, we can see that
the final output state $\ket{\phi_j}$ is exactly $\ket{\phi}$.

As an illustrative example,  let us consider the special case when
$n=2$. Without loss of generality, we can assume $\ket{0'}=\ket{0}$
and $\ket{1'}=\ket{1}$. In this case there are only two permutations
$I$ and $X$. Then theorem \ref{schmidtsep} indicates that
$\ket{\psi}$ can be transformed to $\ket{\phi}$ if and only if
$\lambda_1\leq \mu_1$. We shall describe a simple protocol for the
transformation. If $\lambda_1=\mu_1$, then $\ket{\psi}$ and
$\ket{\phi}$ are equivalent up to some local unitary. Assume that
$\lambda_1<\mu_1$. We can choose $0<p<1$ such that
\begin{equation}
[p I+(1-p)X]\mu=\lambda.
\end{equation}
A simple calculation indicates that
$$p=\frac{\lambda_1-\mu_2}{\mu_1-\mu_2}.$$
Then Alice performs a measurement $\{M_0,M_1\}$ on her subsystem,
where
\begin{equation}
M_0=\sqrt{p}(\sqrt{\frac{\mu_0}{\lambda_0}}\ket{0}\bra{0}+\sqrt{\frac{\mu_1}{\lambda_1}}\ket{1}\bra{1})
\end{equation}
and
\begin{equation}\
M_1=\sqrt{1-p}(\sqrt{\frac{\mu_1}{\lambda_0}}\ket{0}\bra{0}+\sqrt{\frac{\mu_0}{\lambda_1}}\ket{1}\bra{1}).
\end{equation}
If the measurement outcome is $0$, then the final output state is
already $\ket{\phi}$ and nothing needs to be done. Otherwise, every
party performs a bit-flip operation $X$ on their subsystems.

We would like to point out that many results valid for bipartite
pure states can be directly generalized to multipartite pure states
with generalized Schmidt decompositions, as these results only
depend on Schmidt coefficient vectors \cite{LO97, JP99a, JP99,
VID99, VJN00, FM00,DK01, SRS02, FWX02, FDY04, DFY05,SDY05,
CS05,DJFY06,AN07,TUR07}. For instance, using the very same method,
we can easily show the following generalized Vidal's formula
\cite{VID99} for a multipartite quantum system.

\begin{theorem}\upshape The maximal conversion probability from $\ket{\psi}$ to
$\ket{\phi}$ using LOCC is given by
\begin{equation}\label{keyeq3}
p_{\max}(\psi,\phi)=\min\{\frac{E_l(\psi)}{E_l(\phi)}:0\leq l\leq
n-1\},
\end{equation}
where $E_l(\psi)=\sum_{k=l}^{n-1} \lambda_k$.
\end{theorem}

The right-hand side of Eq. (\ref{keyeq3}) is an upper bound for
$p_{\max}$ and follows directly from the optimality of Vidal's
result for bipartite pure states, as we can always split all parties
into two groups as we did in the proof of theorem 1. We only need to
show that there really exists a local protocol which can attain this
upper bound. Our protocol is almost the same as the one given by
Vidal \cite{VID99}. We first convert $\ket{\psi}$ into a temporary
state $\ket{\Omega}$ with certainty, then further convert
$\ket{\Omega}$ into $\ket{\phi}$ by performing a local measurement.

More precisely, suppose $\gamma=(\gamma_0,\ldots,\gamma_{n-1})$ is
the Schmidt coefficient vector of $\ket{\Omega}$ which is
constructed using the same method as in Ref. \cite{VID99}. From the
construction of $\ket{\Omega}$, we obtain that $\lambda\prec\gamma$.
Applying theorem 1, we can make sure the conversion from
$\ket{\psi}$ to $\ket{\Omega}$ can be done using LOCC with
certainty. So the first step in Vidal's proof to convert
$\ket{\psi}$ into $\ket{\varphi}$ can be achieved locally. The proof
that $\ket{\Omega}$ can be locally transformed into $\ket{\phi}$
with probability $p_{\max}$ is exactly the same as that in Ref.
\cite{VID99}.

As another instance, one can easily see that surprising phenomena
such as entanglement catalysis \cite{JP99}, partial recovery of
entanglement \cite{FM00}, multiple-copy entanglement transformation
\cite{SRS02}, and mutual catalysis \cite{FWX02} also exist in
multipartite quantum systems, and many interesting properties
concerning with these phenomena are still valid for multipartite
pure states with generalized Schmidt decompositions
\cite{DK01,FDY04,DFY05,SDY05,CS05,DJFY06,AN07,TUR07}.

We are indebted to the colleagues in the Quantum Computation and
Quantum Information Research Group for many enjoyable conversations.
In particular, we sincerely thank Professor Mingsheng Ying for his
numerous encouragement and constant support of this research, and
one of us (R. D.) wishes to thank Professor Yuan Feng for countless
inspiring discussions on entanglement transformation. This work was
partly supported by the National Natural Science Foundation of China
(Grants Nos. 60621062 and 60503001) and the Hi-Tech Research and
Development Program of China (863 project) (Grant No. 2006AA01Z102).

\end{document}